\documentclass[a4paper,aps,11pt,preprint,showpacs]{revtex4}
\usepackage{graphicx,amsmath,revsymb,bm,amssymb,amsfonts,amsthm}
\begin{document}

\title{Nonrelativistic double photoeffect on lithiumlike ions at high
energies}
\author{A.I.~Mikhailov,$^{1,2}$ I.A.~Mikhailov,$^{1}$ A.V.~Nefiodov,$^{1,3}$
and G.~Plunien$^3$}
\affiliation{$^1$Petersburg Nuclear Physics Institute, 188300 Gatchina,
St.~Petersburg, Russia \\
$^2$Max-Planck-Institut f\"ur Physik komplexer Systeme,
N\"othnitzer Stra{\ss}e 38, D-01187 Dresden, Germany \\
$^3$Institut f\"ur Theoretische Physik, Technische Universit\"at
Dresden, Mommsenstra{\ss}e 13, D-01062  Dresden, Germany }

\date{Received \today}
\widetext
\begin{abstract}
The total cross section for double ionization of lithiumlike ions by
a high-energy photon is calculated in leading order of the
nonrelativistic perturbation theory. The partial contributions due
to simultaneous and sequential emissions of two electrons are taken
into account. The cross section under consideration is shown to be
related to those for double photoeffect on the ground and excited
$2{}^{1,3}S$ states of heliumlike ions. The double-to-single
ionization ratio is equal to $R = 0.288/Z^{2}$ for lithiumlike ions
with moderate nuclear charge numbers $Z$. However, even for the
lightest three-electron targets such as Li and Be$^+$, analytical
predictions are found to be in good agreement with the numerical
calculations performed within the framework of different rather
involved approaches.
\end{abstract}
\maketitle

Recent progress in developments of intense sources of the
synchrotron radiation has raised considerable interest in the
theoretical and experimental investigations of the double ionization
of atoms and ions following by the absorption of a single photon.
The double photoionization is a process of fundamental importance,
because it is entirely caused by the electron correlations. As a
target, it became usual to chose neutral helium or heliumlike ions,
which represent the simplest multi-electron systems. The extensive
studies of the double photoeffect on helium atom in the ground state
allowed one to deduce detailed information about the inner-shell
electron correlations \cite{1,2,3}.

The growing number of theoretical calculations has been recently
devoted to the double photoeffect on excited metastable $2{}^{1,3}S$
states of the helium isoelectronic sequence
\cite{4,5,6,7,8,9,10,11,12,13}. The problem is also of particular
interest, because it is concerned with the study of inter-shell
electron correlations. However, due to lack of intense sources of
the metastable helium, there has been no experimental work on the
topic. In addition, the direct measurements on heliumlike ions with
moderate values of nuclear charge number $Z \gg 1$ are hampered in
this case due to absence of long-living excited states.

Another direction of present experimental and theoretical
investigations is focused on the double and even triple
photoionization of three-electron targets in the ground state
\cite{14,15,16,17,18,19,20,21,22}. At present, the lightest targets,
namely, the lithium atom or the Be${}^+$ ion, are used only. For
lithium, the double-to-single ionization ratio
$R=\sigma^{++}/\sigma^+$ has been measured for photon energies
$\omega < 1$ keV, which is still essentially below the high-energy
nonrelativistic asymptotic limit. After reaching the maximum value
of about $4.6\%$ around $\omega \simeq 240$ eV, the ratio $R$
declines slowly up to $3.9\%$ at $\omega \simeq 910$ eV \cite{17}.

In the high-energy nonrelativistic limit, the theoretical
calculations of double and triple photoionizations of Li have been
first performed with the use of B-spline basis sets \cite{19}. The
asymptotic double-to-single ionization ratio is predicted to be
$3.37(3)\%$, which is consistent with the available experimental
data. The high-energy limits for the ratio $R$ were also calculated
within the framework of the multiconfigurational Hartree-Fock method
\cite{20} and with the use of fully correlated variational wave
functions \cite{21}. The value of $3.36\%$ reported by Yan \cite{21}
is in perfect agreement with the calculation performed by van der
Hart and Greene \cite{19}, while the ratio of $1.81\%$ predicted by
Cooper \cite{20} is significantly lower. In the work \cite{21}, the
calculation is also extended to the ground state of Be${}^+$ ion,
which yields $\sigma^{++}/\sigma^+ =1.97\%$. Recently, the double
and triple photoionization of Li have been calculated for the
near-threshold energy range, employing the time-dependent
close-coupling approach \cite{22}.

In contrast to the sophisticated numerical methods mentioned above,
we shall present an analytical evaluation of the total cross section
for double photoionization of lithiumlike ions within the
high-energy nonrelativistic limit. The cross section under
investigation is also shown to be related to those for double
photoeffect on the ground and excited $2{}^{1,3}S$ states of
heliumlike ions with the same nuclear charge number $Z$.
Accordingly, experiments with stable three-electron targets can
allow one to test the theoretical predictions for low-lying excited
states of heliumlike ions.

The present study is performed for asymptotic photon energies
$\omega$ characterized by $I \ll \omega \ll m$, where $I=m(\alpha
Z)^2/2$ is the binding energy of the K-shell electron, $m$ is the
electron mass, and $\alpha$ is the fine-structure constant
($\hbar=1$, $c=1$). The ejected electrons are considered as being
nonrelativistic. Accordingly, the Coulomb parameter is supposed to
be sufficiently small, that is, $\alpha Z \ll 1$. Since the photon
interacts with a single electron, the emission of two electrons is
mediated only via the electron-electron interaction. The latter is
taken into account within the framework of perturbation theory. As a
zeroth approximation, Coulomb wave functions are employed. Although
the formal parameter of the perturbation theory is $1/Z$, the actual
expansion turns out to converge extremely fast for any value of the
nuclear charge number $Z$ \cite{23}. Accordingly, the present
consideration applies to the lithium isoelectronic sequence within
the range $3 \leqslant Z \lesssim 30$.

In contrast to the heliumlike targets, the double photoionization of
lithiumlike ions can proceed via three channels, namely, double
ionization of both K-shell electrons, double ionization of K- and
L-shell electrons, and single ionization of one K-shell electron
with excitation of the other K-shell electron. In the latter case,
the remaining ion results to be in the doubly excited state, which
can decay afterwards either due to the Auger effect or due to
radiative emission. For multicharged ions with not too large values
of $Z$, the Auger decay accompanied by electron ejection is the
dominant process. The first two channels represent the simultaneous
(direct) ionization, while the third channel can be referred to as
the sequential (indirect) ionization. In a real experiment, one
usually measures the number of ions, the charge of which has been
increased by two units. Therefore, all three channels contribute to
the total cross section of double ionization of lithiumlike ions
\begin{equation}
\sigma^{++}(\mathrm{Li}) =\sigma_{\mathrm{sim}}^{++}(\mathrm{Li}) +
\sigma_{\mathrm{seq}}^{++}(\mathrm{Li}) \, .
\label{eq1}                                        
\end{equation}

Let us first consider the direct photoionization. Our calculations
are performed in leading order of the nonrelativistic perturbation
theory with respect to the electron-electron interaction. In this
approximation, one has to consider only the Feynman diagrams with
one-photon exchange. Accordingly, it is sufficient to take into
account the interaction only between two active electrons, which are
involved in real transitions. The interaction with the third
electron (spectator) is neglected, since it can first contribute
only in the next-to-leading order of the perturbation theory.
Therefore, the cross section
$\sigma^{++}_{\mathrm{KK}}({\mathrm{Li}})$  for double ionization of
the K-shell electrons in lithiumlike ion is equal to the cross
section $\sigma^{++}({\mathrm{He}})$ for double photoeffect on
heliumlike ion in the ground state. The latter can be cast into the
following analytical form \cite{24,25}
\begin{equation}
\sigma^{++}(\mathrm{He}) =\sigma^{+}(\mathrm{He})Z^{-2} B_1 \, .
\label{eq2}                                        
\end{equation}
Here $B_1=0.090$ and $\sigma^{+}(\mathrm{He})$ is the total cross
section for single photoeffect on heliumlike ion in the high-energy
limit. Within the approximations employed here, one can write
\begin{equation}
\sigma^{+}(\mathrm{He}) =  2\sigma^{+}_{\mathrm{K}} \,  ,
\label{eq3}
\end{equation}
where
\begin{equation}
\sigma^{+}_{\mathrm{K}}  = \frac{2^7 \pi \alpha}{3 m \omega}
\left(\frac{I}{\omega}\right)^{5/2} \,
\label{eq4}
\end{equation}
is the total cross section for single photoionization of the K-shell
electron in the Born high-energy limit \cite{26}.

\begin{figure}[h]
\centerline{\includegraphics[scale=0.6]{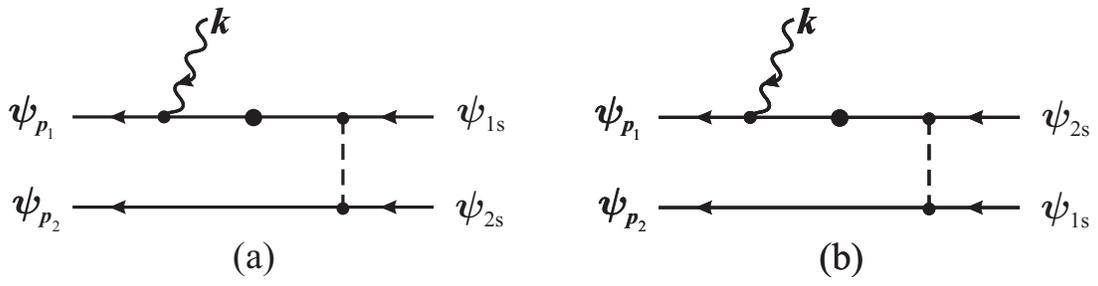}}
\caption{\label{fig1} Feynman diagrams for ionization of the K- and
L-shell electrons following by the absorption of a single photon.
Solid lines denote electrons in the Coulomb field of the nucleus,
dashed line denotes the electron-electron Coulomb interaction, and
the wavy line denotes an incident photon. The line with a heavy dot
corresponds to the Coulomb Green's function.}
\end{figure}

Now we shall consider in more details the simultaneous ejection of
two electrons from different shells (K and L) following by the
absorption of a single photon. In the high-energy nonrelativistic
limit, the double photoionization is known to proceed mainly due to
the electron-electron interaction in the initial state, providing
the Coulomb gauge is employed \cite{24}. Another contribution to the
amplitude of the process, which is due to the electron-electron
interaction in the final state, turns out to be of about the factor
$I/\omega$ smaller and, therefore, can be neglected. In addition, if
$\omega \gg I$, the photon energy is distributed among the ejected
electrons extremely nonuniformly \cite{24}. The main contribution to
the cross section arises from the edge domains of the electron
energy spectrum, where the energy of one electron is much larger
than that of the second electron, that is, either $E_{p_1} \sim
\omega$ and $E_{p_2} \sim I$ or $E_{p_1} \sim I$ and $E_{p_2} \sim
\omega$. In the following, we shall label the fast and slow
electrons by the indices 1 and 2, respectively. Then there is just
one edge domain characterized by the restriction $p_1 \gg p_2$ on
the momenta of escaping electrons at infinity. Accordingly, one
needs to take into account only the Feynman diagrams depicted in
Fig.~\ref{fig1}. The total amplitude of the process is given by
\begin{equation}
M =M_a  - M_b \, ,
\label{eq5}                                                
\end{equation}
where
\begin{eqnarray}
M^{\vphantom{*}}_a &=& w^*_{\lambda_1}w^*_{\lambda_2}
{\mathcal A}^{\vphantom{*}}_a \,
w^{\vphantom{*}}_{\mu_1} w^{\vphantom{*}}_{\mu_2} \, ,
\label{eq6}   \\
M^{\vphantom{*}}_b &=&w^*_{\lambda_1}w^*_{\lambda_2} {\mathcal
A}^{\vphantom{*}}_b \, w^{\vphantom{*}}_{\mu_2}
w^{\vphantom{*}}_{\mu_1} \, ,
\label{eq7}       \\
{\mathcal A}_a &=&\langle \psi_{\bm{p}_1} \psi_{\bm{p}_2} |
\hat{V}_\gamma G_C(E)V_{12} | \psi_{1s} \psi_{2s} \rangle  \,  ,
\label{eq8} \\
{\mathcal A}_b&=&\langle \psi_{\bm{p}_1} \psi_{\bm{p}_2}
|\hat{V}_\gamma G_C(E) V_{12} | \psi_{2s} \psi_{1s} \rangle \,  .
\label{eq9}
\end{eqnarray}
Here $w_\mu$ denotes the Pauli spinor characterized by the spin
projection $\mu$ relative to a quantization axis, $G_C(E)$ is the
nonrelativistic Coulomb Green's function with the energy $E=E_{1s} +
E_{2s}- E_{p_2}$,  $E_{p_2}$ is the energy of slow electron in the
continuous spectrum, and $E_{1s}$ and $E_{2s}$ are the
single-electron energies of the orbitals $1s$ and $2s$,
respectively. In the coordinate representation, the operators
$\hat{V}_\gamma$ and $V_{12}$, which describe the electron-photon and
electron-electron interactions, read
\begin{eqnarray}
\hat{V}_\gamma &=&  -\frac{i}{m}\frac{\sqrt{4\pi \alpha}}
{\sqrt{2\omega}} \, e^{i(\bm{k} \cdot \bm{r})}  \left(\bm{\mathrm e}
\cdot \bm{\nabla} \right) \, ,
\label{eq10}\\
V_{12} &=& \frac{\alpha}{|\bm{r}_1 - \bm{r}_2|}  \,  ,
\label{eq11}
\end{eqnarray}
where $\bm{r}_1$ and  $\bm{r}_2$ are the electron coordinates,
$\bm{k}$ is the photon momentum, and $\omega = |\bm{k}| = k$ and
$\bm{\mathrm e}$ are the energy and the polarization vector of a
photon, respectively. The latter obeys the conditions $(\bm{\mathrm
e}\cdot \bm{k}) = 0$ and $(\bm{\mathrm e}^*\cdot \bm{\mathrm e}) =
1$. Since the operators \eqref{eq10} and \eqref{eq11}, and the
nonrelativistic Green's function do not contain the spin matrices,
the amplitudes \eqref{eq6} and \eqref{eq7} take the forms
\begin{eqnarray}
M_a &=& {\mathcal A}_a \delta_{\lambda_1 \mu_1}
\delta_{\lambda_2 \mu_2} \,  ,
\label{eq12}  \\
M_b &=& {\mathcal A}_b \delta_{\lambda_1 \mu_2} \delta_{\lambda_2
\mu_1} \,  .
\label{eq13}
\end{eqnarray}
Analytical expressions for the amplitudes \eqref{eq8} and
\eqref{eq9} have been obtained in Ref.~\cite{7}. In the high-energy
limit, both amplitudes are real.

The fivefold differential cross section for double photoionization
of the KL electron pair of lithiumlike ion is given by
\begin{eqnarray}
d^5\sigma^{++}_{\mathrm{KL}}(\mathrm{Li}) &=&  \frac24
\sum_{\lambda_1, \lambda_2}
\sum_{\mu_1, \mu_2} | M |^2 d^5\Gamma  \, ,
\label{eq14}   \\
d^5\Gamma &=& m p_1 \frac{d\Omega_1}{(2 \pi)^2}\frac{d^3\bm{p}_2}{(2
\pi)^3} \,  ,
\end{eqnarray}
where $d\Omega_1$ is the solid angle of fast electrons and the
amplitude $M$ is defined by Eqs.~\eqref{eq5}, \eqref{eq12}, and
\eqref{eq13}. The expression \eqref{eq14} involves summation over
the electron polarizations in the final state and averaging over
those of the initial state. The factor 2 accounts for the existence
of two K-shell electrons; each of them can be coupled with the
L-shell electron. Inserting the amplitude $M$ in Eq.~\eqref{eq14},
one obtains
\begin{equation}
d^5\sigma^{++}_{\mathrm{KL}}(\mathrm{Li}) = 2 ( {\mathcal A}^2_a  +
{\mathcal A}^2_b - {\mathcal A}_a  {\mathcal A}_b ) \,d^5\Gamma \, .
\label{eq16}
\end{equation}

According to the results obtained in the work \cite{7}, the cross
sections for double photoionization of heliumlike ions in the $2^1S$
and $2^3S$ states can be also expressed through the same amplitudes
${\mathcal A}_a$ and ${\mathcal A}_b$ as follows
\begin{eqnarray}
d^5\sigma^{++}_{\mathrm{s}}(\mathrm{He}^*) &=&  ( {\mathcal A}^2_a
+ {\mathcal A}^2_b + 2 {\mathcal A}_a  {\mathcal A}_b )
\,d^5\Gamma  \,  ,
\label{eq17}  \\
d^5\sigma^{++}_{\mathrm{t}}(\mathrm{He}^*) &=&  ( {\mathcal A}^2_a +
{\mathcal A}^2_b - 2 {\mathcal A}_a  {\mathcal A}_b ) \,d^5\Gamma
\,  .
\label{eq18}
\end{eqnarray}
The subscripts "$\mathrm{s}$`` and "$\mathrm{t}$`` refer to the
singlet and triplet states, respectively, while the notation
$\mathrm{He}^*$ implies the heliumlike ion in the excited $1s2s$
state. Taking into account the statistical weights for the singlet
and triplet states, one obtains the averaged cross section for
double photoeffect on the metastable heliumlike ion
\begin{equation}
d^5\sigma^{++}(\mathrm{He}^*) = \frac14 \,
d^5\sigma^{++}_{\mathrm{s}}(\mathrm{He}^*) + \frac34
\,d^5\sigma^{++}_{\mathrm{t}}(\mathrm{He}^*) = ( {\mathcal A}^2_a  +
{\mathcal A}^2_b - {\mathcal A}_a {\mathcal A}_b ) \,d^5\Gamma  \, .
\label{eq19}
\end{equation}
Comparing Eqs.~\eqref{eq16} and \eqref{eq19}, one finds a simple
relation between the cross sections of KL-ionization of Li- and
He$^*$-like ions
\begin{equation}
d^5\sigma^{++}_{\mathrm{KL}}(\mathrm{Li}) = 2\,
d^5\sigma^{++}(\mathrm{He}^*)   \,  .
\label{eq20}
\end{equation}

Adding the expression \eqref{eq20} to the partial contribution for
the direct KK-ionization, one obtains the differential cross section
for the simultaneous double photoionization of lithiumlike ion
\begin{equation}
d^5\sigma^{++}_{\mathrm{sim}}(\mathrm{Li}) =
d^5\sigma^{++}_{\mathrm{KK}}(\mathrm{Li}) +
d^5\sigma^{++}_{\mathrm{KL}}(\mathrm{Li}) =
d^5\sigma^{++}(\mathrm{He}) + 2\,d^5\sigma^{++}(\mathrm{He}^*) \, .
\label{eq21}
\end{equation}

Integrating Eq.~\eqref{eq21} over the solid angles of electron
ejections and averaging it over the photon polarizations, one
arrives at the energy distribution of slow electrons
\begin{eqnarray}
d\sigma^{++}_{\mathrm{sim}}(\mathrm{Li}) &=&
d\sigma^{++}(\mathrm{He}) + 2\, d\sigma^{++}(\mathrm{He}^*)=
\sigma^{+}(\mathrm{Li}) Z^{-2} \beta(\varepsilon_2) d\varepsilon_2
\,  ,  \label{eq22} \\
\beta(\varepsilon_2) &=& \frac{16}{17}\left\{Q_1(\varepsilon_2) +
\frac14 \, K_{\mathrm{s}}(\varepsilon_2)+ \frac34 \,
K_{\mathrm{t}}(\varepsilon_2) \right\} \,  ,
\label{eq23}
\end{eqnarray}
where $\varepsilon_2 =E_{p_2}/I$ denotes the dimensionless energy of
the low-energy electron. The explicit expression for the function
$Q_1(\varepsilon_2)$ reads \cite{25}
\begin{equation}
Q_1(\varepsilon_2) = \frac{8 J^2(\varepsilon_2)}{1 - \exp (-2
\pi/\sqrt{\varepsilon_2} )} \,
\end{equation}
together with
\begin{eqnarray}
J(\varepsilon_2) &=& \frac{8 \zeta^2}{(1 + \zeta)^3} \left\{
\frac{J_1}{\varepsilon_2 + 1} - \frac{J_2}{\varepsilon_2 + 2}
\right\} \,  , \qquad \zeta = (\varepsilon_2 +
2)^{-1/2} \,  ,   \\
J_1 &=& \exp\left(- \frac{2}{\sqrt{\varepsilon_2}}
\arctan \sqrt{\varepsilon_2}\right) \int_0^1
\frac{t^{-\zeta} (1 -t)}{(1 + q t)^3}\, dt \, ,  \\
J_2 &=& \int_0^1 \frac{t^{-\zeta} (1 -t)^3}{(1 + q t)^3}\,
\Phi_1(t) \Phi_2(t) \, dt \,  , \qquad
q = \frac{1 -\zeta}{1 + \zeta} \,  ,
\end{eqnarray}
where
\begin{eqnarray}
\Phi_1(t) &=& \exp\left(-\frac{2}{\sqrt{\varepsilon_2}} \arctan
\frac{\sqrt{\varepsilon_2} (1 - t)}{ a +bt } \right) \, ,\qquad a=
\sqrt{\varepsilon_2 +2} +2\,  ,  \\
\Phi_2(t) &=& \frac{(3\zeta^2 +1)(1-t)^2 + 6\zeta(1-t^2) +
2(1 + t)^2}{
[(2\zeta^2 +1)(1-t)^2 + 4 \zeta (1-t^2) + (1 +t)^2]^2 } \,  ,
\qquad b= \sqrt{\varepsilon_2 +2} - 2\,  .
\end{eqnarray}

In Eq.~\eqref{eq23}, the functions $K_{\mathrm{s,t}}(\varepsilon_2)
= (9/8) R_{\mathrm{s,t}}(\varepsilon_2)$ are also expressed via the
definite integrals over the elementary functions \cite{7}
\begin{eqnarray}
R_{\mathrm{s,t}}(\varepsilon_2) &=& \frac{2^7}{9} \biggl\{\lambda
\int_1^\infty \left(\frac{y+1}{y-1} \right)^{1/\lambda} \bigl[
\chi_1(y) \pm \chi_2(y) \bigr] \, dy \biggr\}^2 \,  , \qquad
\lambda=\sqrt{\varepsilon_2 + 5/4} \,   ,
\label{eq30} \\
\chi_1(y) &=& ( x - 1/2)^{-3}\bigl\{ \varphi(1/2 ,1) +
\varphi(1/2,2)/2 - (x^2 - 2x + 5/4 )
\varphi(x,2) +\nonumber\\
&+&  2 ( x - 1/2)^2 (x -1)\varphi(x,3) -
\varphi(x,1) \bigr\} \,  ,  \\
\chi_2(y) &=& (x-1)^{-4}(x -5/2)\bigl\{ \varphi(1 ,1) -
\varphi(x,1) - (x -1)^2 \varphi(x,2) \bigr\} + \nonumber\\
&+& (2x - 1)(x-1)^{-1} \varphi(x ,3) \,   , \\
\varphi(\nu,k) &=& ( \nu^2 + \varepsilon_2)^{-k}
\exp\left(-\frac{2}{\sqrt{\varepsilon_2}} \arctan
\frac{\sqrt{\varepsilon_2}}{\nu} \right) \,  ,
\qquad x=\lambda\,y +3/2 \,   .
\end{eqnarray}
In Eq.~\eqref{eq30}, the signs "$+$`` and "$-$`` correspond to the
indices "${\mathrm{s}}$`` and "${\mathrm{t}}$``, respectively.

Within our approximations, the cross section
$\sigma^{+}(\mathrm{Li})$ for single photoeffect on lithiumlike ions
in the ground state is related to those on hydrogenlike and
heliumlike ions as follows
\begin{equation}
\sigma^{+}(\mathrm{Li}) =2 \sigma^{+}_\mathrm{K} +
\sigma^{+}_\mathrm{L} = \frac{17}{8}\, \sigma^{+}_\mathrm{K} =
\frac{17}{16}\,\sigma^{+}(\mathrm{He}) \,  .
\label{eq34}
\end{equation}
In addition, the cross section $\sigma^{+}(\mathrm{He^*})$ for
single ionization of heliumlike ions in the excited $1s2s$ state is
given by
\begin{equation}
\sigma^{+}(\mathrm{He^*}) = \sigma^{+}_\mathrm{K} +
\sigma^{+}_\mathrm{L} = \frac{9}{8}\, \sigma^{+}_\mathrm{K}  \,  .
\label{eq35}
\end{equation}

\begin{figure}[h]
\centerline{\includegraphics[scale=1.2]{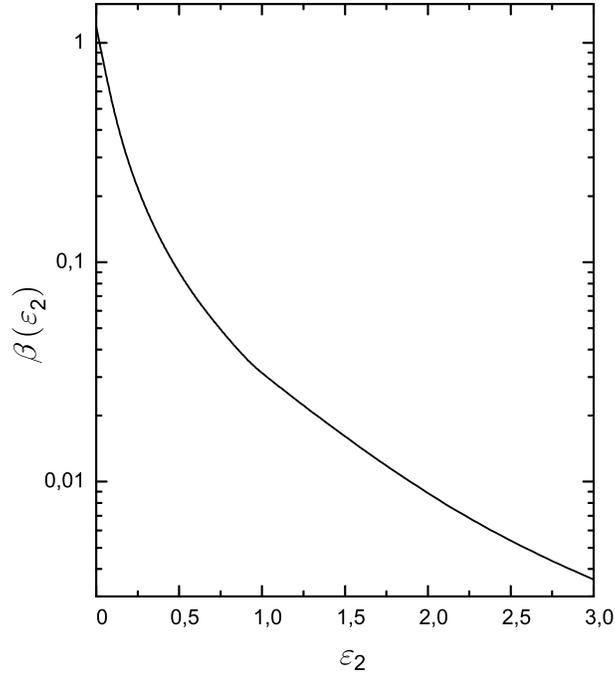}}
\caption{\label{fig2} The function $\beta(\varepsilon_2)$ is
calculated with respect to the dimensionless energy $\varepsilon_2$
of the slow electron according to  Eq.~\protect{\eqref{eq23}}. }
\end{figure}

The resulting energy function $\beta(\varepsilon_2)$ defined by
Eq.~\eqref{eq23} is depicted in Fig.~\ref{fig2}. Integrating
Eq.~\eqref{eq22} over the energy $\varepsilon_2$ of the slow
electron yields the total cross section for the direct double
photoionization
\begin{eqnarray}
\sigma^{++}_{\mathrm{sim}}(\mathrm{Li}) &=& \sigma^{++}(\mathrm{He})
+ 2\,\sigma^{++}(\mathrm{He}^*)= \sigma^{+}(\mathrm{Li}) Z^{-2}B
\,   ,  \label{eq36}\\
B&=&\frac{16}{17}\left\{B_1 + \frac98 \left[  \frac14 B_{\mathrm{s}}
+ \frac34 \, B_{\mathrm{t}} \right] \right\} \,   ,
\label{eq37}
\end{eqnarray}
where
\begin{equation}
B_1 = \int_0^\infty  Q_1(\varepsilon_2) \, d\varepsilon_2 \,  ,
\qquad B_{\mathrm{s,t}} = \int_0^\infty
R_{\mathrm{s,t}}(\varepsilon_2) \,d\varepsilon_2 \,  .
\label{eq38}
\end{equation}
In general, the upper limit in the integrals \eqref{eq38} is the
finite number $\varepsilon_\gamma/2$, where $\varepsilon_\gamma
=\omega/I$ is the dimensionless energy of the incoming photon. It
should be chosen in such a way to fulfil the condition
$\varepsilon_\gamma \gg \varepsilon_2$. However, since the
integrands decrease very rapidly for increasing $\varepsilon_2$, the
upper limit can be trend to infinity. Then the numerical values for
the constants \eqref{eq38} are equal to $B_1 =0.090$ \cite{24,25},
$B_{\mathrm{s}} = 0.318$ \cite{7}, and $B_{\mathrm{t}}=0.0575$
\cite{7}. In fact, the energy integrals are already saturated at
$\varepsilon_2 \simeq 1$. In view of the relation \eqref{eq37}, one
finally obtains $B=0.215$. This is our first result.

Now we shall consider the indirect double photoionization of
lithiumlike ions in the ground state. For light multicharged ions,
the corresponding cross section
$\sigma_{\mathrm{seq}}^{++}(\mathrm{Li})$ coincides with that
$\sigma_{\mathrm{KK}}^{+*}(\mathrm{Li})$ for single ionization with
excitation. In this case, one K-shell electron leaves the ion, while
the another one undergoes a transition into a higher-lying state
forming together with the $2s$-electron the doubly excited states,
which afterwards decay via emission of the Auger electron. To
leading order of the nonrelativistic perturbation theory, electron
excitations can occur only into the $ns$ states with the principal
quantum numbers $n \geqslant 2$ \cite{27}. Since the $2s$ state is
occupied just by half in Li-like ions and it is vacant in He-like
ions, the cross section $\sigma_{\mathrm{KK}}^{+*}(\mathrm{Li},2s)$
for the single ionization of lithiumlike ions with excitation into
the $2s$ state is half as large as that
$\sigma^{+*}(\mathrm{He},2s)$ for heliumlike ions. In the case of
transitions into other excited states, the cross sections for
three-electron systems are equal to those for two-electron ions.
Accordingly, one can write down the following chain of relations
\begin{equation}
\sigma_{\mathrm{seq}}^{++}(\mathrm{Li}) =
\sigma_{\mathrm{KK}}^{+*}(\mathrm{Li}) = \sum_{n \geqslant 2}
\sigma_{\mathrm{KK}}^{+*}(\mathrm{Li},ns) = \frac12 \,
\sigma^{+*}(\mathrm{He},2s) + \sum_{n \geqslant 3}
\sigma^{+*}(\mathrm{He},ns)   \, .
\label{eq39}
\end{equation}

To leading order of the perturbation theory with respect to the
electron-electron interaction, the cross sections
$\sigma^{+*}(\mathrm{He},ns)$ for the single ionization of
heliumlike ions with excitation into the $ns$ states read
\cite{27}
\begin{eqnarray}
\sigma^{+*}(\mathrm{He},ns) &=& \sigma^{+}(\mathrm{He}) Z^{-2} Q(n)
\,    ,  \label{eq40} \\
Q(n) &=& \frac{2^8}{n^3} \biggl\{ \varkappa \int_1^\infty
\left(\frac{y+1}{y-1} \right)^{1/\varkappa}  \chi(y)(1 + \varkappa
y)^{-3} dy \biggr\}^2  \, ,
\label{eq41} \\
\chi(y) &=& \phi(1,1) - \phi(v,1) - (v-1)^2 \phi(v,2)\,  ,
\qquad \varkappa =\sqrt{2 - n^{-2}} \, ,
\label{eq42} \\
\phi(\nu,k)&=& \frac{(\nu -n^{-1})^{n-k}}{(\nu + n^{-1})^{n+k}} \, ,
\qquad v=2 + \varkappa y \,  .
 \label{eq43}
\end{eqnarray}

For the lowest principal quantum numbers $n$, the numerical values
of the function \eqref{eq41} are equal to $Q(2)=9.11 \cdot 10^{-2}$, 
$Q(3)=1.71 \cdot 10^{-2}$,  $Q(4)=0.62 \cdot 10^{-2}$, 
$Q(5)=0.30 \cdot 10^{-2}$,  $Q(6)=0.17 \cdot 10^{-2}$, 
$Q(7)=0.10 \cdot 10^{-2}$,  $Q(8)=0.7 \cdot 10^{-3}$, 
$Q(9)=0.5 \cdot 10^{-3}$.  In the asymptotic limit $n \gg 1$, 
the following relation
\begin{equation}
n^3 Q(n)|_{n \gg 1} \simeq 0.334
\end{equation}
holds.

Inserting Eq.~\eqref{eq40} into Eq.~\eqref{eq39} and taking into
account the relation \eqref{eq34} yields as our second result
\begin{eqnarray}
\sigma_{\mathrm{seq}}^{++}(\mathrm{Li}) &=& \sigma^{+}(\mathrm{Li})
Z^{-2} C \,  ,
\label{eq45}\\
C &=& \frac{16}{17} \biggl\{ \frac12 Q(2) + \sum_{n \geqslant 3}
Q(n) \biggr\}  \,   .
\end{eqnarray}
Employing the numerical data for the function $Q(n)$, we obtain
$C=0.073$.

The double-to-single photoionization ratio $R$ for lithiumlike ions,
which is usually measured experimentally, is given by
\begin{equation}
R= \frac{\sigma^{++}(\mathrm{Li})}{\sigma^{+}(\mathrm{Li})} =
\frac{B+C}{Z^2} = \frac{0.288}{Z^2}\,  ,
\label{eq47}
\end{equation}
where Eqs.~\eqref{eq1}, \eqref{eq36}, and \eqref{eq45} have been
employed. As follows from the ratio $C/B \simeq 1/3$, about one
third of the total cross section for double photoeffect is gained on
account of the channel of sequential ionization. The universal
scaling law \eqref{eq47} is our third result. The highest absolute
accuracy for the universal curve can be expected for lithiumlike
multicharged ions with moderate nuclear charge numbers $10 \lesssim
Z \lesssim 25$, since the higher-order corrections omitted in the
present calculations are of minor importance for the total cross
sections. Nevertheless, for the particular cases of the neutral
lithium and the Be$^+$ ion, our predictions for the ratios $R$ are
equal to $3.2\%$ and $1.8\%$, respectively, which are already in
good agreement with the numerical results obtained within the
framework of rather complicated methods in works \cite{19,21}.

Concluding, we have established the universal high-energy behavior
of the double-to-single photoionization ratio for lithiumlike ions
in the ground state. The partial contributions due to the direct and
indirect ionization channels are taken into account. We have also
found relations between the cross sections for double photoeffect on
Li- and He-like ions with the same nuclear charge number $Z$. These
relations can be employed for experimental tests of theoretical
predictions concerning the double photoionization of low-lying
excited states of He-like ions.

\acknowledgements

A.M. is grateful to the Dresden University of Technology for the
hospitality and for financial support from Max Planck Institute for
the Physics of Complex Systems. A.N. and G.P. acknowledge financial
support from DFG, BMBF, and GSI. This research was also supported in
part by RFBR (Grant no. 05-02-16914) and INTAS (Grant no.
03-54-3604).

\newpage

\end{document}